\documentclass[twocolumn,showpacs,preprintnumbers,amsmath,amssymb]{revtex4}
\usepackage{graphicx}
\usepackage{dcolumn}
\usepackage{bm}
\begin{document}
\preprint{APS/123-QED}
\title{THE MECHANICS OF THE SYSTEMS OF STRUCTURED PARTICLES }
\author{V.M. Somsikov}
 \altaffiliation[] {}
 \email{nes@kaznet.kz}
\affiliation{%
Laboratory of Physics of the geoheliocosmic relation, Institute of
Ionosphere, Almaty, Kazakstan.
}%

\date{\today}
\begin{abstract}

The mechanics of the structured particles develops. The
substantiation of applicability of such mechanics for the
description of processes of evolution in open nonequilibrium
systems is offered. The consequences following from the equations
of dynamics of structured particles are analyzed.
\end{abstract}

\maketitle

\section{\label{sec:level1}Introduction\protect}

All the natural systems are opened and nonequilibrium. The openness
of systems is defined by character of power interrelation with an
external space. The level of nonequilibrium is defined by the
deviation of entropy of system from its maximum value. If openness
and nonequilibrium are small and their influence on the studying
phenomenon is slightly then that phenomenon can be described within
the frame of classical mechanics. But the using of classical
mechanics meets with the big difficulties if one analyzes the
mechanisms of structures creations or the transitive and nonlinear
phenomena, i.e. all those type of dynamics which are connected with
dissipation and an openness of systems. The most vivid example of
such difficulties is the irreversibility problem. Since the L.
Boltzmann the interest to this problem is not decreasing [1-4].

The property of mixing of the Hamilton's systems and averaging of
phase space on physically small volume are used in the basis of
the existing explanations of irreversibility. The explanation of
the nature of averaging is based on postulation of casual
fluctuations [2, 3]. Thus the casualty is the irreversibility
reason. But it means absence of determinism of the nature. On
other side if the world is cognizable, the deterministic mechanism
of irreversibility exists. Our research have shown that such
mechanism exists in the frameworks of the expanding of the
classical mechanics. The expansion consists in replacement of
model of system in the form of set of material points on model of
system in the form of the set of structured particles [14]. The
dynamics of such system can be described with the help of the
motion equation of the structured particles consisting of
potentially interacting material points or equilibrium subsystems
(ES). Such model possesses the big generality because the wide
range of nonequilibrium systems can be presented in the form of ES
set.

The analysis of the submitted model of the nonequilibrium systems
is constructed under the following conditions: 1). Energy of ES
should be presented as the sum of internal energy and energy of ES
motion; 2). The each element of system should belong to one ES;
3). The ES are in equilibrium during all time.

The first condition is necessary for introduction the internal
energy into the description of systems dynamics as the key
parameter characterizing the energy exchanges between ES. The
second condition allows to avoid the difficulties arising due to
mixing of particles between different ES. Last condition used in
thermodynamics. It is equivalent to the condition of the weak
enough ES interaction which does not disturb the equilibrium.

The task of this work is to show how and why the possibility of
the description of nonequilibrium systems are appearing as a
result of replacement the model of system consisting of the
material points by model of system which consists of the
structural particles.

For this purpose we will determine restrictions of model of system
from material points which do not allow studying of nonequilibrium
systems and we will show why replacement of this model by model in
the form of ES set gives the possibility to remove these
restrictions. We will also show how it is possible to obtain the
motion equation for ES and how Lagrange, Hamilton, Liouville
equations for ES follow from it. We will explain interrelation of
the classical mechanics with thermodynamics and how it is possible
to define entropy in the classical mechanics.

\section{The approach substantiation.}
In works on an irreversibility explanation the models of hard
bodies under condition of unacceptable for classical mechanics
postulation of presence of fluctuations are used [2,3]. It is
possible to exclude necessity of fluctuations postulation if one
considers the system which consists of the set of ES instead of
system model of separate hard bodies. In this case the process of
equilibration can be connected with the work of forces between ES
which transform the ES motion energy into their internal energy
[5-7]. I.e. representation of system in the form of ES opens the
possibility to explain irreversibility within the frame of the
classical mechanics.

Firstly, the models in the form the set of ES were used by Gibbs
for creation of the statistical physics [8]. The same models are
used in the kinetic methods for analyzing of the nonequilibrium
systems [13].

In the description of nonequilibrium systems on the basis set of
ES it is necessary to take into account their interaction. Really,
nonequilibrium systems are characterized by the presence of the
dissipative structures. They are created and supported by streams
of energy, substance and the entropy caused by a field of forces
in system. The ES distribution function is defined by the two
parameters: their internal energy and their motion energy [13,
17].

Let us notice that the representation of energy of system in the
form of the sum of internal energy and the energy of its motion is
used for solution of a problem of two bodies. It is made by
transition to the system of the center of mass (CM). In the
laboratory system of co-ordinates the two-body problem can not be
solved due to the nonlinearity caused by influence of motion of
one body on another.

These statements lead us to assumption: the description of
dynamics of nonequilibrium system in the classical mechanics is
possible if this system will be represented as a set of ES
consisting of potentially interacting material points and the
energy of ES will be presented as the sum of the motion energy and
internal energy. Basing on such model it is possible to obtain the
equation of ES interaction from the law of energy conservation.
These equations will connect the microstreams of energy caused by
pair interactions of particles with macrostreams of energy between
ES.

\section{The equation of ES motion }
The motion equations for two ES can be obtained in two stages. At
first we obtain the motion equation for the system of material
points in the non-homogeneity space. After that we obtain the
motion equations for two interacting ES. Forces between ES can be
obtained from their potential energy of interaction.

Let us firstly show how the motion equation for a system of $N$
material points (further we will call it particles) with weights
$m=1$ can be obtained [14,18]. Forces between pairs of particles
are central and potential. The systems energy $E$ is equal to the
sum of kinetic energy $T_N=\sum\limits_{i=1}^{N} m{v_i}^2/2$,
their potential energy in a field of external forces, -
${U_N}^{env}$, and the potential energy of their interaction
${U_N}(r_{ij})={\sum\limits_{i=1}^{N-1}}{\sum\limits_{j=i+1}^{N}}U_{ij}(r_{ij})
$, where $r_{ij}=r_i-r_j$, $r_i, v_i$ are the coordinate and
velocity of $i$-th particle. Thus,
$E=E_N+U^{env}=T_N+U_N+U^{env}=const$.

By transition to corresponding variables we submitted the systems
energy as a sum of CM motion energy and internal energy. After
derivation this energy on time, we will obtain [14]:
\begin{eqnarray}
V_NM_N\dot{V}_N+{\dot
E}_N^{ins}=-V_NF^{env}-\Phi^{env}\label{eqn1}
\end{eqnarray}
Here $F^{env}=\sum\limits_{i=1}^{N}F_i^{env}(R_N,\tilde{r}_i)$,
${\dot E}_N^{ins}={\dot T}_N^{ins}(\tilde{v}_i)+{\dot
U}_N^{ins}(\tilde{r}_i)$=
$\sum\limits_{i=1}^{N}\tilde{v}_i(m\dot{\tilde{v}}_i+F(\tilde{r})_i)$,
 $\Phi^{env}=\sum\limits_{i=1}^{N}\tilde{v}_iF_i^{env}(R_N,\tilde{r}_i)$,
$r_i=R_N+\tilde{r}_i$, $M_N=mN$, $v_i=V_N+\tilde{v}_i$,
$F_i^{env}=\partial{U^{env}}/\partial{\tilde{r}_i)}$,
$\tilde{r}_i$, $\tilde{v}_i$ are the coordinate and velocity of
$i$-th particle in relative to the system CM, $R_N,V_N$ are the
coordinate and velocity of the system's CM.

The eq. (1) represents balance of energy of system in a field of
external forces. The first term in the left hand side determines
change of kinetic energy of system -
${\dot{T}}_N^{tr}=V_NM_N\dot{V}_N$. The second term determines the
change of internal energy of system, ${\dot{E}}_N^{ins}$.

Because $\sum\limits_{i=1}^{N}\tilde{v}_i=0$, the change of
internal energy will be distinct from zero only when the
characteristic scale of inhomogeneity of an external field is
commensurable with system scale. In this case depending on a
configuration of an external field can vary or a kinetic energy of
system rotation or energy of the relative motion of elements. In
the both cases the force changing an internal energy is
non-potential.

Let us compare dynamics of a particle with the dynamics of their
system. As it follows from the Newton equation {\it{the particle
motion is defined by the work of potential forces which transform
the energy of an external field into the kinetic energy}}.

The work of external forces for a system goes both on change,
$T_N^{tr}$ and on change of $E_N^{ins}$. I.e. in similar to the
energy, the external force breaks up on two parts. The first part
is potential force. It changes the systems momentum. The second
forces is non-potential. It changes the internal energy. The work
of these forces is not equal to zero when the forces for different
particles are different. Hence, {\it{the system motion is defined
by work of potential and non-potential forces which transform the
energy of an external field into kinetic energy of the system
motion and into internal energy accordingly}}.

By multiplying the eq. (1) on $V_N$ and dividing on $V_N^2$ we
found the equation of system motion [18]:
\begin{eqnarray}
M_N\dot{V}_N= -F^{env}-{\alpha_N}V_N\label{eqn2}
\end{eqnarray}

where $\alpha_N=[{\dot E}_N^{ins}+\Phi^{env}]/V_N^2$  is a
coefficient which determine the change of internal energy.

Let us call eq. (2) as generalized Newton equation (GNE) for the
structured particle. The first term in the right hand side defines
system acceleration, and the second term defines change of its
internal energy. The GNE is reduced to the Newton equation if one
neglects the relative motion of elements, i.e. when the internal
energy does not change. In this case the dynamics of system is
similar to the reversible dynamics of an elementary particle.

The eqs. (1,2) can be obtained also by multiplying the Newton's
equation on corresponding velocity and then summarizing all
equations [15]. But if we simply summarize Newton's equations the
non-potential forces will be lost and the right hand side will be
equal to the potential forces. It is a Newton equation for material
point with the weight equal to the sum of weights of all particles.
Thus the Newton's equation does not define full dynamics of system
since it does not include the non-potential forces. It confirms
Leibnitz idea that vis viva, i.e. energy is a fundamental parameter
but not momentum or force [10].

Thus the system dynamics in an external field is defined by two
parameters: the motion energy and internal energy. The change of
the motion energy is caused by potential force; the change of
internal energy is caused by the non-potential force.

Let us explain how to obtain the ES interaction equations. For
this purpose we take the system consisting of two ES-$L$ and $K$.
The $L$-is a number of elements in the $L$-ES and $K$-in $K$-ES,
i.e. $L+K=N$. Let $LV_L+KV_K=0$, where $V_L$ and $V_K$ are
velocities of $L$ and $K$-ES. By derivating energy of system on
time, we will obtain:
${\sum\limits_{i=1}^{N}v_i{\dot{v}}_i}+{\sum\limits_{i=1}^{N-1}}\sum\limits_{j=i+1}^{N}v_{ij}
F_{ij}=0$, where $F_{ij}=U_{ij}=\partial{U}/\partial{r_{ij}}$.

For finding the equation for $L$-ES,  we gather at the left hand
side only the terms defining the change of kinetic and potential
energy of interaction of $L$-ES elements among themselves. All other
terms we displaced into the right hand side and combined the groups
of terms in such a way that each group contained of the terms with
identical velocities. In accordance with Newton equation, the groups
which contain terms with velocities of the elements from $K$-ES are
equal to zero. As a result the right hand side of the equation will
contain only the terms which determine the interaction of the
elements $L$-ES with the elements $K$-ES. Thus we will have:
${\sum\limits_{i_L=1}^{L}}v_{i_L}
{\dot{v}}_{i_L}+{\sum\limits_{i_L=1}^{L-1}}\sum\limits_{j_L=i_L+1}^{L}
F_{{i_L}{j_L}}v_{{i_L}{j_L}}={\sum\limits_{i_L=1}^{L}}\sum\limits_{j_K=1}^{K}
F_{{i_L}{j_K}}v_{j_K}$ where double indexes are entered for a
designation of an accessory of a particle to corresponding
subsystem. If we will make replacement
$v_{i_L}=\tilde{v}_{i_L}+V_L$, where $\tilde{v}_{i_L}$ is a velocity
of $i_L$ particle in relative to CM of $L$ -ES then we obtain the
equation for $L$-ES. The equation for $K$-ES can be obtained in the
same way. As a result we will have [9, 14]:
\begin{eqnarray}
V_LM_L\dot{V}_L+{\dot{E}_L}^{ins}=-{\Phi}_L-V_L{\Psi}
\end{eqnarray}
\begin{eqnarray}
V_KM_K\dot{V}_K+{\dot{E}_K}^{ins}={\Phi}_K+V_K{\Psi}
\end{eqnarray}

Here $M_L=mL, M_K=mK, \Psi=\sum\limits_{{i_L}=1}^LF^K_{i_L}$;
${\Phi}_L=\sum\limits_{{i_L}=1}^L\tilde{v}_{i_L}F^K_{i_L}$;
${\Phi}_K=\sum\limits_{{i_K}=1}^K\tilde{v}_{i_K}F^L_{i_K}$;
$F^K_{i_L}=\sum\limits_{{j_K}=1}^KF_{i_Lj_K}$;
$F^L_{j_K}=\sum\limits_{{i_L}=1}^LF_{i_Lj_K}$;
${\dot{E}_L}^{ins}={\sum\limits_{i_L=1}^{L-1}}\sum\limits_{j_L=i_L+1}^{L}v_{i_Lj_L}
[\frac{{m\dot{v}}_{i_Lj_L}}{L}+\nonumber\\+F_{i_Lj_L}]$;
${\dot{E}_K}^{ins}={\sum\limits_{i_K=1}^{K-1}}\sum\limits_{j_K=i_K+1}^{K}v_{i_Kj_K}
[\frac{{m\dot{v}}_{i_Kj_K}}{K}+\nonumber\\+F_{i_Kj_K}]$.

The eqs. (3,4) are the ES interaction equations. They describe
energy exchange between ES. Independent variables of ES are
macroparameters and microparameters. Macroparameters are
coordinates and velocities of ES motion. Microparameters are
coordinates and velocities of ES elements.

The ES binds together two types of the description: on the
macrolevel and on the microlevel. The description on the macrolevel
determines of ES dynamics as a whole and on the microlevel
determines dynamics of ES elements.

The potential force, $\Psi$, determines the motion of ES as a
whole. This force is the sum of the potential forces acting on
elements of one ES from another.

The non-potential forces which determined by the terms ${\Phi}_L$
and ${\Phi}_K$, will transform the motion energy of ES into the
internal energy as a result of chaotic motion of elements one ES
in the field of the forces of another ES. They are dependent on
velocities and cannot be expressed as a gradient from any scalar
function. These forces are equivalents to dissipative forces. As
well as in case of system in an external field these forces are
distinct from zero when the characteristic scale of inhomogeneity
of a field of forces of one ES is commensurable with scale of
another.

The ES motion equations corresponding to the eqs. (3,4) can be
writen [14]:
\begin{equation}
M_L\dot{V}_L=-\Psi-{\alpha}_LV_L \label{eqn5}
\end{equation}
\begin{equation}
M_K\dot{V}_K=\Psi+{\alpha}_KV_K\label{eqn6}
\end{equation}
where ${\alpha}_{L}=(\dot{E}^{ins}_{L}+{\Phi}_{L})/V^2_{L}$,
${\alpha}_{K}=({\Phi}_{K}-\dot{E}^{ins}_{K})/V^2_{K}$,

The eqs. (5, 6) are GNE for ES. The second terms in the right hand
side of the equations determine the force changing internal energy
of ES. This force is equivalent to the friction force. The
efficiency of transformation of energy of relative motion of ES
into internal energy are determined by the factors "$\alpha_L$",
"$\alpha_K$". If the relative velocities of ES elements are equal
to zero the force of friction is also equal to zero.

When the change of internal energy can be neglected, the GNE will be
transformed into the Newton equation for material points. For
example it is possible when distances between ES are great enough
[16].

The eqs. (1, 5, 6) allow to define the non-potential forces in the
nonequilibrium system.

\section{The generals of Lagrange, Hamilton and Liouville equations for ES}

The Hamilton principle for material points is deduced from
differential D'Alambert principle with the help of the equation of
Newton [10-12]. For this purpose the integral on time from the
virtual work made by effective forces is equated to a zero.
Integration on time is carried out provided that external forces
possess power function. It means that the canonical principle of
Hamilton is fair only for cases when $\sum F_i\delta R_i=-\delta
U$ where $i$ is a number of particles, and $F_i$ - is a force
acting on this particle. But for ES it is impossible to demand
performance of a condition of conservatism of forces because the
non-potential forces exist. Non-potential forces change ES
internal energy. Therefore Lagrange, Hamilton, Liouville equations
for structural particles must be deduced basing on ES motion
equations [6, 7].

Liouville equation for ES looks like [6, 7]:
\begin{equation}
df/dt=-\sum\limits_{L=1}^{R}{\partial}{F_L}/{\partial}V_L
\label{eqn7}
\end{equation}

Here $f$-is a distribution function for a set of ES, $F_L$-is a
force acting on $L$-ES, $L=1,2,3...R$ is a number of ES, $V_L$-is
a velocity of $L$-ES. These forces can be found with the help of
eqs. (5, 6).

The right hand side of the eq. (7) is not equal to zero since forces
of ES interaction depend on velocities of elements.

The state of system as a set of ES can be determined by the point in
the phase space which consists of $6R-1$ coordinates and momentums
of ES, where $R$ is a number of ES. Let us call this space as
$S$-space to distinguish it from usual phase space for material
points. The $S$-space unlike usual phase space is compressible
though total energy of all elements is a constant. The rate of
compression of $S$-space is determined by the rate of transformation
of motion energy of the ES into their internal energy. Thus the
volume of compression of $S$-space is determined by energy of the ES
motion.

The impossibility of return of internal energy of ES in its energy
of motion is caused by impossibility of change of momentum ES due to
the motion of its material points. Formally it follows from
independence of variables coordinates and velocities for ES and for
their material points [11]. Therefore the system will aspire to
equilibrium.

\section{The equations of interaction of systems and thermodynamics}

Let us consider how thermodynamics can follow from the classical
mechanics [14, 18]. According to the basic equation of
thermodynamics the work of external forces acting on the system
are splitting on two parts. The first part is connected to
reversible work. The change of the motion energy of system as
whole can be put in conformity for this energy part. The second
part of energy will go on heating. It is connected with the
internal degrees of freedom of system. The internal energy of ES
corresponds to this part of energy.

Let us take the motionless nonequilibrium system consisting of "$R$"
of ES. Each of ES consists of great number of elements $N_L>>1$,
where $L=1,2,3...R, N=\sum\limits_{L=1}^{R}N_L$. Let $dE$ is the
change of the energy of a system (do not confuse $E$ with the
internal energy of ES - $E^{ins}$). It is known from the
thermodynamics: ${dE=dQ-PdY}$. Here, according to common
terminology, $E$ is the energy of a system; $Q$ is the thermal
energy; $P$ is the pressure; $Y$ is the volume. The equation of a
systems interaction also includes two types of energy. The one part
goes to the change of ES motion. The other part changes the internal
energy. The interrelation of classical mechanics and thermodynamics
in more details is considered in [14].

Entropy can be entered into the classical mechanics as the rate
characterizing increasing of the internal energy ES at the expense
of energy of their motion. Then the entropy increasing will be
defined so [9, 14]:
\begin{equation}
{{\Delta{S}}={\sum\limits_{L=1}^R{\{{N_L}
\sum\limits_{k=1}^{N_L}\int[{\sum\limits_s{{F^{L}_{ks}}v_k}/{E^{L}}]{dt}}\}}}}\label{eqn8}
\end{equation}

Here ${E^{L}}$ is the kinetic energy of $L$-ES; $N_L$ is the
number of elements in $L$-ES; $L=1,2,3...R$; ${R}$ is the number
of ES; ${s}$ is a number of the external elements which interact
with elements ${k}$ belonging to the $L$-ES; ${F_{ks}^{L}}$ is a
force, acted on $k$-element; $v_k$ -is a velocity of the $k$-
element.

The expression for the entropy production and definition of the
necessary conditions for a stationary state of the nonequilibrium
system are following from the accepted definition of entropy [18].

\section{Conclusion}
The obtained results lead to the following conclusions. System
evolution in non-homogeneity space is determined by external force.
The external force breaks up on potential and non-potential parts.
The motion energy of the system changes by the potential component
of the force. The system internal energy changes by the work of
non-potential part. This work is distinct from zero if the scale of
heterogeneity of external forces is less or commensurable with the
system scale.

Evolution of the closed nonequilibrium system in the homogeneous
space which represented by a set of ES, as well as in the case of
a system motion in an external field, is defined by potential and
non-potential forces. But the difference is following. These
forces are caused by interaction between ES instead of external
forces. The potential part of forces between ES changes their
kinetic motion energy. The work of non-potential part of the force
transforms the energy of ES motion into the internal energy.
Therefore the phase space which defined in coordinates and the
velocities of ES CM is compressible. We call that space as
$S$-space. $S$-space compression is defined by Liouville equation
for ES. The system equilibrates when the ES motion energy
transforms into their internal energy. It defines the
irreversibility mechanism.

The existence of the non-potential forces in nonequilibrium
systems throws light on the nature of non-integrability of
Hamilton systems [2, 16]. Really, the self agreement between
changes of potential and kinetic energy of particles would exist
if the forces were only potential. And it would mean systems
integrability or possibility of its description by means of
Newton's equation. But Newton's equation for material points
excludes non-potential forces. Therefore it does not allow to
consider all streams of energy in nonequilibrium system. Hence it
does not allow to describe nonequilibrium systems with the help of
Hamilton formalism.

For the same reasons there are difficulties of the description of
strong interactions by means of an initial formalism of Hamilton.
These difficulties can be overcome using the modified equations
for case when potential interactions of elements inside ES are
more strong than interactions between ES as, for example, in the
case of interaction of elementary particles. Then it will be
possible to take into account the energy of excitation of internal
degrees of freedom of particles and change of their internal
energy. It is one of possible reasons of broken symmetry.

The equations of ES motion help us to include the frictional
forces into the classical mechanics strictly. Thereby
substantiation of the including of a weak dissipation in the
regions of Hamilton's system resonances is justified. It is
important for development of the theory of the deterministic chaos
lying in the basis of physics of evolutionary processes.

As a whole, the obtained equations connect the classical mechanics
and thermodynamics.  The explanation of the First law of
thermodynamics is based on the fact that the work of subsystems'
interaction forces changes both the energy of their motion and
their internal energy. The explanation of the Second law of
thermodynamics is connected with irreversible transformation of
the subsystems' relative motion energy into their internal energy.
Moreover the impossibility of occurrence of unstructured particles
in the classical mechanics follows from it. It is equivalent to
the infinite divisibility of the matter.

Thus many difficulties of the description of nonequilibrium
systems within the limits of the classical mechanics can be
overcome by replacement of model of material points by models of
the systems consisting from ES, by transition from Newton's
equation to GNE and by using of a formalism of Hamilton for ES.

\smallskip


\medskip
\begin{thebibliography}{9}

\bibitem{Ref1}
Cohen E.G, Boltzmann and statistical mechanics, Dynamics: Models
and Kinetic Methods for Nonequilibrium Many Body systems.1998 NATO
Sci. Series E: Applied Sci., 371, p. 223

\bibitem{Ref2}
Prigogine I, From the being to becoming. 1980, Moscow

\bibitem{Ref3}
Zaslavsky G.M, Chaotic dynamic and the origin of Statistical
laws,1999, Physics Today, August, Part 1, p.39

\bibitem{Ref4}
 Ulenbek G., Fundamental problems of the stat. mechanics, 1971, UFN.
V.103, N27, p.275

\bibitem{Ref5}
Somsikov, V.M. Non-recurrence problems in evolution of a hard-disk
system. Int. Jour. Bifurc. And Chaos. 2001, 11, p.2863

\bibitem{Ref6}
Somsikov V.M, Some approach to the Analysis of the Open
Nonequilibrium systems, 2002, AIP, 20, p.149

\bibitem{Ref7}
Somsikov V.M, Equilibration of a hard-disks system. Intern. Jour.
Bifurc. and Chaos, 2004, 14, 11, p.4027

\bibitem{Ref8}
Landau L.D,  Lifshits Ye. M., Statistical Physics. 1976, Nauka,
Moscow

\bibitem{Ref9}
 Somsikov V.M., Thermodynamics and classical mechanics. Journal of
physics. Conference series, 2005, 23, p.7

\bibitem{Ref10}
 Lanczos C., The variation principles of mechanics. 1962, Univer. of
Toronto press

\bibitem{Ref11}
 Landau L.D,  Lifshits Ye.M., Mechanics. 1958, Nauka, Moscow

\bibitem{Ref12}
Goldstein G, Classical mechanics.1975, Moscow

\bibitem{Ref13}
Rumer Yu.B, Ryvkin M.Sh, Thermodynamics. Stat. Physics and
Kinematics. 1977, Nauka, Moscow

\bibitem{Ref14}
Somsikov V.M, Expansion of a formalism of classical mechanics for
nonequilibrium systems. arX:physics/0703141v1 17Mar2007.

\bibitem{Ref15}
Longmair K, Plasma physics. 1966, Atomizdat, Moscow

\bibitem{Ref16}
Poincare A., About science. Ì., 1983, Nauka, Moscow

\bibitem{Ref17}
Klimontovich, Yu.L. Stat. physics. 1982, Nauka, Moscow

\bibitem{Ref18}
V.M. Somsikov. The restrictions of classical mechanics in the
description of dynamics of nonequilibrium systems and the way to get
rid of them.arXiv:0805.1186v1 [physics.class-ph] 8 May 2008.


\end{thebibliography}
\end{document}